\documentclass[useAMS,usenatbib,asm]{mn2e}
\usepackage{graphicx,fleqn,times}
  
\title{Can bars be destroyed by a central mass concentration? 
  I~Simulations} 
\author[E.~Athanassoula, J.~C.~Lambert, W.~Dehnen]
       {E.~Athanassoula$^{1}$, J.~C.~Lambert$^{1}$, W.~Dehnen$^{2}$\\
$^1$ Observatoire de Marseille, 
2 Place Le Verrier, 
F-13248 Marseille Cedex 4, France \\
$^{2}$ Dept. of Physics and Astronomy, University of Leicester,
  Leicester, LEI 7RH\\
}

\date{Accepted .
      Received ;
      }

\pagerange{\pageref{firstpage}--\pageref{lastpage}}
\pubyear{2005}

\begin{document}

\maketitle

\label{firstpage} 
\begin{abstract}
  We study the effect of a central mass concentration (CMC) on the
  secular evolution of a barred disc galaxy. Unlike previous studies,
  we use fully self-consistent 3D $N$-body simulations with live
  haloes, which are known to be important for bar evolution. The CMC
  is introduced gradually, to avoid transients. In all cases where the
  mass of the CMC is of the order of, or more than, a few per cent of
  the mass of the disc, the strength of the bar decreases noticeably.
  The amount of this decrease depends strongly on the bar type. For
  the same CMC, bars with exponential surface-density profile, which
  formed in a disk-dominated galaxy (MD-type bars), can be totally
  destroyed, while 
  strong bars with a flat surface-density profile, whose evolution is
  largely due to the halo (MH-type bars), witness only a decrease of
  their strength. 
  This decrease occurs simultaneously from both the innermost and
  outermost parts of the bar. The CMC has a stronger effect on the
  Fourier components of higher azimuthal wave number $m$, leading to
  fatter and/or less rectangular bars.  Furthermore, the CMC changes
  the side-on outline from peanut-shaped to boxy or, for massive CMCs,
  to elliptical. Similarly, side-on initially boxy outlines can be destroyed.
  The CMC also influences the velocity dispersion profiles.  Most of
  the decrease of the bar strength occurs while the mass of the CMC
  increases and it is accompanied by an increase of the pattern speed.
  In all our simulations, the mass of the CMC
  necessary in order  to destroy the bar is at least several per cent
  of the mass of the disc. This argues that observed
  super-massive black holes are not likely to destroy pre-existing bars.
\end{abstract}


\section{Introduction}
Both bars and central mass concentrations (hereafter CMCs) are
features that are present in the majority of disc galaxies. Thus, bars
harbouring a CMC should be the rule rather than the exception and it
is important to study the effects of the latter on the evolution of
the former (and vice versa). Observations in near infrared
wavelengths, where the effect of dust is largely suppressed, reveal
that between 70\% and 90\% of all disc galaxies have a bar component
\citep*{SeigarJames1998, EskridgeEtal2000, GrosbolPatsisPompei2004}.
CMCs -- such as massive black holes, central discs, dense central
stellar clusters -- are very frequent, too.

Super-massive black holes have a mass of the order of
$10^6-10^9\,M_{\odot}$, which correlates with the luminosity, the
mass, and, in particular, the velocity dispersion of the hosting bulge
\citep*{KormendyRichstone1995, MagorrianEtal1998,
  FerrareseMerritt2000, GebhardtEtal2000, TremaineEtal2002,
  McLureDunlop2002, MarconiHunt2003}. Such correlations lead to a
ratio of super-massive black hole mass to bulge mass of the order of
$10^{-3}$ \citep[see also][]{KormendyGebhardt2001}. The mass of the
CMC is in fact larger than that of the black hole, since the black
hole pulls inwards the bulge and disc material in its immediate
neighbourhood and thus forms a cusp, which contributes to the CMC mass
\citep*{Peebles1972, ShapiroLightman1976, BahcallWolf1976, Young1980,
  GoodmanBinney1984, QuinlanHernquistSigurdsson1995,
  LeeuwinAthanassoula2000}.

Compact massive central discs, stellar and/or gaseous, are also
examples of CMCs. Gaseous such discs are expected to form particularly
frequently in barred galaxies, due to the gas inflow driven by the bar
\citep*{Athanassoula1992b, WadaHabe1992, WadaHabe1995, FriedliBenz1993,
  HellerShlosman1994, ReganTeuben2004}. Indeed, observations have
shown that large concentrations of molecular gas can be found in many
disc galaxies, and particularly in barred ones
\citep{SakamotoEtal1999, ReganEtal2001, HelferEtal2003}. The mass of
such components is of the order of $10^7-10^9\,M_{\odot}$, while their
radial extent varies from several tens of parsecs to a couple of kpc.
Discy bulges \citep{Athanassoula2005b} are examples of mainly stellar
such discs \citep[][and references therein]{KormendyKennicutt2004}.
CMCs can also be central dense stellar clusters, with masses of
$10^6-10^8\,M_{\odot}$.

Orbital-structure studies, aimed at understanding the effect of the
CMC on the bar-supporting orbits, were made first in 2D
\citep{HasanNorman1990} and then extended to 3D
\citep*{HasanPfennigerNorman1993}. They include the calculation of the
main families of periodic orbits and of their stability. The latter is
particularly important, since stable periodic orbits trap around them
regular orbits, while unstable periodic orbits generate chaos.
These studies showed that with
increasing CMC mass or concentration the unstable fraction of the
bar-supporting orbits increases, reducing the possibility for barred
equilibria. 

Orbital-structure studies have thus made the first crucial step in our
understanding of the effect of a CMC on a bar by revealing the
mechanism by which a CMC could destroy a bar. These studies can inform
us whether a given bar model (i.e.\ density profile and pattern speed)
may co-exist with a given CMC. However, the parameter space of bar
models is huge and an extensive exploration is neither possible nor
meaningful, because nothing will be learned about the \emph{evolution}
of a given bar in presence of a growing CMC. 

Such information can only be achieved by appropriate $N$-body
simulations, which were pioneered by
\citeauthor*{NormanSellwoodHasan1996} (\citeyear[][ hereafter NSH]
{NormanSellwoodHasan1996}). These authors used particles on a grid to
represent the (barred) disc, which was exposed to the gravitational
field of the CMC and of a spheroidal (bulge-like) component. For both 2D
and 3D simulations, NSH found that a CMC of mass 5\% of that of the
galaxy is sufficient for bar dissolution. In the $N$-body simulations of
\citeauthor{ShenSellwood2004} (\citeyear[][ hereafter
SS]{ShenSellwood2004}) the disc is represented by particles on a 3D
cylindrical polar grid and, similarly to NSH, exposed to the gravity
of the CMC and of a rigid spherical halo component. These authors
found that a
CMC mass of a few percent is necessary to destroy the bar, by and large,
consistent with the results of NSH. More precisely, in the models
displayed in their Fig.~5, a CMC mass of approximately 4\% that of the
disc was necessary for a hard (compact) CMC and more than 10\% for a
soft (extended) CMC. They compared two models with different bar
strengths and concluded that the strength of the bar hardly influences
these numbers.

Recently, \citeauthor{HozumiHernquist1999} (\citeyear[][ hereafter
HH]{HozumiHernquist1998,HozumiHernquist1999,HozumiHernquist2005})
re-examined the problem, 
solving the Poisson equation by expanding the density and the
potential in a set of bi-orthonormal basis functions. Their approach
is 2D, which implies the necessity for a rigid halo potential, similar
to NSH \& SS. These authors find that a CMC of mass 0.5\% of the disc
mass is sufficient to destroy the bar. Thus, their limiting CMC mass
is an order of magnitude lower than that given by NSH. The difference
between their results and those of NSH\,\&\,SS is surely significant.
Indeed, if the mass necessary to destroy a bar is as small as claimed
by HH, then the masses of CMCs observed in disc galaxies are
sufficient to dissolve observed bars in relatively short time scales.
The opposite is true if the NSH \& SS picture is correct.

This disagreement between previous studies, as well as their lack of
realism in modeling the galactic halo, prompted us to revisit the
problem. HH argued that the differences between the two sets of
simulations are due to the different initial density distributions in
the disc, since they use an exponential disc, while NSH\,\&\,SS
employed a Kuzmin-Toomre disc \citep{Kuzmin1956,Toomre1963}. HH argue
that, since the latter is more centrally concentrated than the former,
the CMC could influence a larger fraction of the disc mass and thus
achieve bar dissolution easier. However, preliminary simulations
\citep{AthanassoulaEtal2003} showed that bars are rather robust, and
so we opted for an exponential disc model which is known to represent
well the observed light distribution of disc galaxies
\citep{Freeman1970}.  This may give a better chance for the CMC to
destroy the bar and allows us to check whether indeed the initial
density distribution in the disc is of crucial importance for the
result. We employ a Poisson solver different from both types
previously used in this problem, namely a tree code described in
section~\ref{subsec:code}.

Unlike the previous studies, we use a live halo in our simulations,
which makes them \emph{fully} self-consistent. This has been shown to
be essential for bar formation in presence of a strong halo component.
Indeed, \citeauthor{Athanassoula2002} (\citeyear[][ hereafter
A02]{Athanassoula2002}) compared two simulations with initially
identical disc components and haloes of identical mass distribution.
In one case the halo was rigid (represented as an imposed spherical
potential) and in the other it was composed of about $10^6$ particles.
The difference between the evolution in the two cases is stunning
(Fig.~1 of A02). In the first case, the disc formed only a minor oval
confined to the centre-most parts of the disc, while in the second
case it grew a strong bar with very realistic observed properties
\citep[][ hereafter AM02]{AthanassoulaMisiriotis2002}. This important
difference is due to the fact that the angular momentum absorption by
the halo is artificially suppressed in one of the two cases, while in
the other it is allowed to freely take place \citep[for a
comprehensive discussion of the effect of the halo on the angular
momentum exchange see][ hereafter A03]{Athanassoula2003}.

The paper is organised as follows. In section~\ref{sec:numerical} we
present our numerical methods and our simulations.
Sections~\ref{sec:before_CMC} and \ref{sec:after_CMC} describe the
evolution before and after the introduction of the CMC, respectively.
We discuss our results and conclude in section~\ref{sec:discussion}.

\section{\boldmath $N$-body simulations}
\label{sec:numerical}
We first follow the formation and evolution of a bar in two
bar-unstable discs. We choose a time after the initial fast bar
formation period and during the phase of slow secular evolution. At
this time we start growing a CMC in the centre of the disc. The
various steps in this work are described in the following subsections.
 
\subsection{Initial conditions}
\label{subsec:initcond}
The models are composed of a disc and a halo component. The disc has
an initial volume density profile
\begin{equation}
  \rho_{\mathrm{d}}(R,z)= \frac{M_{\mathrm{d}}}{4 \pi R_d^2 z_0}\;
  \exp (- R/R_{\mathrm{d}})\;\mathrm{sech}^2 (z/z_0),
\end{equation}
where $R$ is the cylindrical radius, $M_{\mathrm{d}}$ is the disc mass
and $R_{\mathrm{d}}$ and $z_0$ are the disc radial and vertical
scale-lengths, respectively\footnote{The system of coordinates is
  taken according to the usual convention that the $x$ and $y$ axes
  are on the disc equatorial plane and the $z$ axis is perpendicular
  to that plane. In all figures the bar will be along the $y$ axis.}.
The initial halo density profile is
\begin{equation} \label{eq:halodens}
  \rho_h (r) = \frac{M_{\mathrm{h}}}{2\pi^{3/2}}\;
  \frac{\alpha}{r_{\mathrm{c}}}
  \;\frac {\exp(-r^2/r_{\mathrm{c}}^2)}{r^2+\gamma^2},
\end{equation}
where $r$ is the radius, $M_{\mathrm{h}}$ is the halo mass and
$\gamma$ and $r_c$ are the halo scale-lengths. The former can be
considered as the core radius of the halo. The constant $\alpha$ is
defined by
\begin{equation}
  \alpha = [1 - \sqrt \pi\,q\,\exp (q^2)\,(1 - \mathrm{erf}(q))]^{-1},
\end{equation} 
where $q=\gamma/r_{\mathrm{c}}$ \citep{Hernquist1993}. Information on
how the initial conditions were set up can be found in
\cite{Hernquist1993} and AM02. Both simulations have
$M_{\mathrm{d}}=1$, $R_{\mathrm{d}}=1$, $z_0=0.2$, $M_{\mathrm{h}}=5$
and $r_{\mathrm{c}}=10$. The first one has a halo with a small core
($\gamma=0.5$) and is thus, following the definition of AM02, MH-type.
The second one has a halo with a large core ($\gamma=5$) and is thus
of MD-type (AM02). The discs of the two simulations have $Q=1.4$ and
1, respectively, where $Q$ is the stability parameter introduced by
\cite{Toomre1964}.

Our model units of mass and length are simply the disc mass and disc
scale length. In order to convert them to $M_\odot$ and kpc, we need
to set the units of length and mass to values representative of the
object in consideration, a barred galaxy. AM02 used a length unit of
3.5 kpc and a mass unit of $5\times10^{10}\,M_{\odot}$. With $G=1$,
this implies that the model unit of velocity is 248\,km\,s$^{-1}$ and
the model unit of time is $1.4\times10^7\,$yr. Thus, time 400
corresponds to $5.6\times10^9\,$yr. This choice, however, is in no way
unique, and we can choose quite different values. For example, we can
use a smaller length unit since the scale-length of the disc increases
with time \citep{ONeillDubinski2003,ValenzuelaKlypin2003} and thus
the initial length unit could be chosen to be 2 kpc. Because of this
rather wide range of possibilities, we will restrict ourselves to
model units and leave it to the readers to convert units according to
their needs.
 
\subsection{\boldmath$N$-body simulations}
\label{subsec:code}
For the simulations described in this paper, we used the publicly
available $N$-body code \textsf{gyrfalcON}, which employs the
tree-based force solver \textsf{falcON} \citep{Dehnen2000:falcON,
  Dehnen2002}. Unlike the standard \cite{BarnesHut1986} tree code,
which has complexity $\mathcal{O}(N\log N)$, \textsf{falcON} uses
cell-cell interactions, has complexity $\mathcal{O}(N)$, and is about
ten times faster at comparable accuracy. Some of the simulations were
run with a proprietary version of the code, which uses the
SSE\footnote{\textbf{S}treaming \textbf{S}IMD \textbf{E}xtensions;
  SIMD = \textbf{S}ingle \textbf{I}nstruction \textbf{M}ultiple
  \textbf{D}ata.} instruction set supported on X86 chips. These
instructions allow a substantial speed-up for the computation of
body-body forces via direct summation.  This method is used within and
between small cells and results in a gain of a factor of $\sim2$ for
this part of the code. The total gain for the force computation for
$N\sim10^6$ bodies (including the tree build) is about 25\%.

In all $N$-body simulations it is necessary to use a softening kernel
in order to avoid excessive noise and diverging forces. A
\cite{Plummer1911} softening, which has density kernel $\propto
(\varepsilon^2 + r^2)^{-5/2}$, has been used in most simulations made
so far. This, however, has the disadvantage that the softened force
converges relatively slowly to the Newtonian force
\citep{AthanassoulaEtal2000, Dehnen2001}, although the difference
becomes negligible, and beyond any numerical importance, at large
radii. Here we adopt a softening with a density kernel
$\propto(\varepsilon^2+r^2)^{-7/2}$ \citep{DehnenEtal2004}, which
converges much faster to the accurate Newtonian force.  This is
particularly interesting for the problem at hand, as will be discussed
in the last section. We adopted a softening length $\varepsilon$ of
0.03, which roughly corresponds to a Plummer softening length of 0.02
(in the sense that the maximum two-body force is the same). A small
softening length is useful in this problem, since the CMC will create
an increased central concentration of the disc and halo components,
and more centrally concentrated structures necessitate a smaller
softening length \citep{AthanassoulaEtal2000}.

The tree code used in our simulations allows for an approximate
interaction (as opposed to the direct summation force) between two
tree nodes (cells or single bodies) only if their critical spheres do
not overlap \citep{Dehnen2000:falcON,Dehnen2002}. The critical spheres
are centred on the centre of mass of the nodes and have radii
$r_{\mathrm{crit}} = r_{\mathrm{max}} / \theta$, where
$r_{\mathrm{max}}$ is the radius containing all particles in the node
and $\theta$ the tolerance parameter (opening angle). In our
simulations we use $\theta=0.5$.

In order to have sufficiently small time steps in the area around the
CMC, we used adaptive time stepping with the block-step scheme (time
steps differ by factors of two and are hierarchically nested). In most
simulations we used 6 levels, in such a way that the longest time step
is $2^{-6}$ and the shortest $2^{-11}$. For the numerically more
demanding simulations (with a CMC of very high mass or very small
radius) we used 7 levels and a smallest time step of $2^{-12}$. The
time step levels are adapted in an (almost) time symmetric fashion to
be on average
\begin{equation}
  \tau = \min\{f_a/a, f_{\Phi}/|\Phi|\},
\end{equation}
where $f_a$ and $f_\Phi$ are constants, $a$ is the modulus of the
acceleration and $\Phi$ is the potential. We adopted $f_a$ = 0.01 and
$f_\Phi$ = 0.015, thus ensuring that the smallest time step bin
contains of the order of a few percent of the particles.

We model the disc with 200\,000 and the halo with roughly 1000\,000 
particles.  With these settings, the
energy conservation after the CMC growth (during CMC growth energy is
not conserved) was always better than one part in 1\,000, and, in most
cases, of the order of a couple of parts in 10\,000.

\subsection{The CMC} \label{subsec:simulations}
Following previous studies (NSH\,\&\,SS), we model the CMC by a Plummer sphere
of potential
\begin{equation}
\Phi_{\mathrm{CMC}} (r) = - \frac {GM (t)}{\sqrt {r^2 + r_{\mathrm{CMC}}^2}},
\end{equation}
where $M(t)$ is the mass of the CMC at time $t$ and $r_{\mathrm{CMC}}$
its radius. In order to avoid transients, the CMC is introduced
gradually with
\begin{equation}
  \frac{M(t)}{M_{\mathrm{CMC}}} = \left\{
    \begin{array} {r@{\quad:\quad}l}
      0 & t \le t_{\mathrm{in}} \\
      \frac {3}{16} \xi^5 - \frac {5}{8} \xi^3 + \frac {15}{16}
      \xi + \frac {1}{2} & t_{\mathrm{in}}<t<t_{\mathrm{grow}}\\
      1 &  t \ge t_{\mathrm{grow}}
    \end{array} \right.
\end{equation}
where $\xi = 2 (t - t_{\mathrm{in}})/t_{\mathrm{grow}} - 1$ and
$t_{\mathrm{in}}$ is the time at which the CMC is introduced
\citep{Dehnen2000:OLR}. Thus the mass, as well as its first and second
derivatives, are continuous functions of $t$.  The radius
$r_{\mathrm{CMC}}$ of the CMC is kept constant with time.  In the
following, since we are interested in the time evolution after the CMC
is introduced, we measure time from the time when the CMC has been
introduced, i.e. from $t_{\mathrm{in}}$.

Some of the software used in the analysis has already been described
in AM02 and we refer the reader to that paper for more information.
Our runs are listed in Table~\ref{tab:initcond_CMC}. The first column
gives the name of the run and the second one the type of the model.
The mass of the CMC, $M_{\mathrm{CMC}}$, its radius,
$r_{\mathrm{CMC}}$, and its growth time, $t_{\mathrm{grow}}$, are
given in the third, fourth and fifth columns, respectively.

\begin{table}
  \caption[]{Parameters for the CMC}
  \begin{flushleft}
    \label{tab:initcond_CMC}
    \begin{tabular}{lllll}
      \hline
      model & initial & $M_{\mathrm{CMC}}$ & $r_{\mathrm{CMC}}$ & 
      $t_{\mathrm{grow}}$ \\
      \hline
      \hline
      MH & MH & 0. & - & -  \\
      MH$_1$ & MH & 0.05 & 0.01 & 100  \\
      MH$_2$ & MH & 0.10 & 0.01 & 100  \\
      MH$_3$ & MH & 0.10 & 0.005 & 100  \\
      MH$_4$ & MH & 0.20 & 0.01 & 100  \\
      MH$_5$ & MH & 0.10 & 0.01 & 200  \\
      \hline
      MD & MD  & 0. & - & -  \\
      MD$_1$ & MD & 0.05 &  0.01 & 100  \\
      MD$_2$ & MD & 0.10 &  0.01 & 100  \\
      MD$_3$ & MD & 0.10 &  0.005 & 100  \\
      MD$_4$ & MD & 0.20 &  0.01 & 100  \\
      MD$_5$ & MD & 0.10 &  0.01 & 200  \\
      MD$_6$ & MD & 0.04 &  0.01 & 100  \\
      MD$_7$ & MD & 0.03 &  0.01 & 100  \\
      MD$_8$ & MD & 0.02 &  0.01 & 100  \\
      MD$_9$ & MD & 0.01 &  0.01 & 100  \\
      MD$_{10}$ & MD & 0.1 &  0.1 & 100  \\
      MD$_{11}$ & MD & 0.1 &  0.08 & 100  \\
      MD$_{12}$ & MD & 0.1 &  0.05 & 100  \\
      MD$_{13}$ & MD & 0.1 &  0.02 & 100  \\
      \hline
    \end{tabular}
  \end{flushleft}
\end{table}

\section{Evolution prior to the introduction of the CMC}
\label{sec:before_CMC}
The formation and evolution of the bar in $N$-body simulations has
already been extensively described in the literature. Here we will
only summarise very briefly some results from AM02, A02, and A03,
which are necessary for discussing the simulations at hand, and will
refer the interested reader to these papers for more information
\citep[see also][]{Athanassoula2005a}.

Barred galaxies evolve by re-distributing their angular momentum. This
is emitted mainly by near-resonant material in the inner disc and
absorbed by near-resonant material in the outer disc and in the halo.
The amount of angular momentum that can be emitted and absorbed
depends on the amount of near-resonant material and on how dynamically
hot/cold this is. Since there is relatively little material in the
outer disc, the halo can play a major role in this exchange, even
though the material that constitutes it is relatively hot. Thus, bars
immersed in massive haloes can grow stronger than bars immersed in
weak haloes, and very much stronger than bars immersed in rigid
haloes. Models where the halo plays a substantial role in the dynamics
within the inner 4 or 5 disc scale lengths, while being also
sufficiently cool to absorb considerable amounts of angular momentum
were termed MH (for Massive Halo) in AM02. Such models form strong
bars. On the contrary, models in which the disc dominates the dynamics
within that region form less strong bars and were termed MD (for
Massive Disc) in AM02. Due to the considerably different dynamics in
the two cases, we decided to study the effect of a CMC in two models,
one of MH-type and one of MD-type.

The observable properties of the bars which grow in these two types of
models are quite different (see AM02 for a more complete discussion).
MH-type bars are stronger (longer, thinner and more massive) than
MD-type bars. Viewed face-on, they have a near-rectangular shape,
while MD-type bars are more elliptical. Viewed
side-on\footnote{The edge-on view in which the bar is viewed
  perpendicular to its major axis is called side-on.The edge-on view
  in which the bar is viewed along its major axis is called end-on. },
they show stronger peanuts and sometimes (particularly towards the end
of the simulation) even `X' shapes. On the other hand, bars in MD-type
models are predominantly boxy when viewed side-on.

\begin{figure}
  \setlength{\unitlength}{2cm}
  \includegraphics[scale=0.65,angle=0]{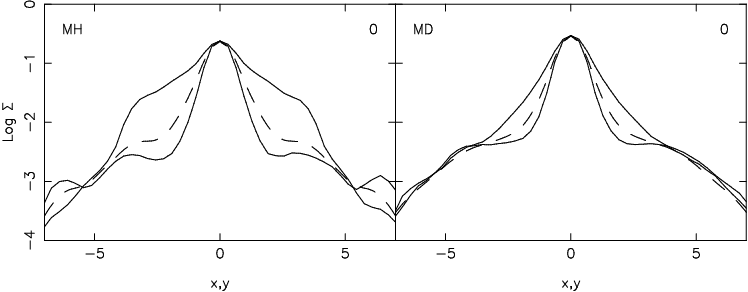}
  \caption{Projected density profiles along the bar major and minor axes
    (solid lines) and azimuthally averaged (dashed lines). The disc is
    seen face-on. The left panel corresponds to model MH and the right
    one to model MD, both at the time the CMC is introduced.}
  \label{fig:faceon_prof}
\end{figure}

Fig.~\ref{fig:faceon_prof} shows the projected density radial profiles
of the two types of bars seen face-on. The results are in good
agreement with what was found by AM02 (see their Fig. 5 and section
5). The radial profile along the major axis of the MD bar decreases
with radius without showing any clear change of slope at the end of the bar.
On the other hand, for the MH type the profile along the bar major
axis is quite different. It has a relatively flat part in the bar
region which is followed by a sharp drop at the end of the bar. These
two different types of profiles have been also found by observations
of barred galaxies. \cite{ElmegreenElmegreen1985}, using blue and
near-infrared surface photometry of a sample of 15 barred galaxies,
classified the photometric profiles in the bar region in two classes:
the flat ones and the exponentially decreasing ones, i.e.\ the two
types found in our two fiducial simulations. The importance of these
two types of profiles for bar destruction will be discussed in the
last section.

\section{Evolution after introduction of the CMC}
\label{sec:after_CMC}
\subsection{The effect of the CMC}
\label{subsec:existence}

\begin{figure*}
  \setlength{\unitlength}{2cm}
  \centerline{MH-type\hspace*{80mm}MD-type}

  \smallskip
  \parbox[t]{70mm}{{}
    \includegraphics[scale=0.4,angle=0]{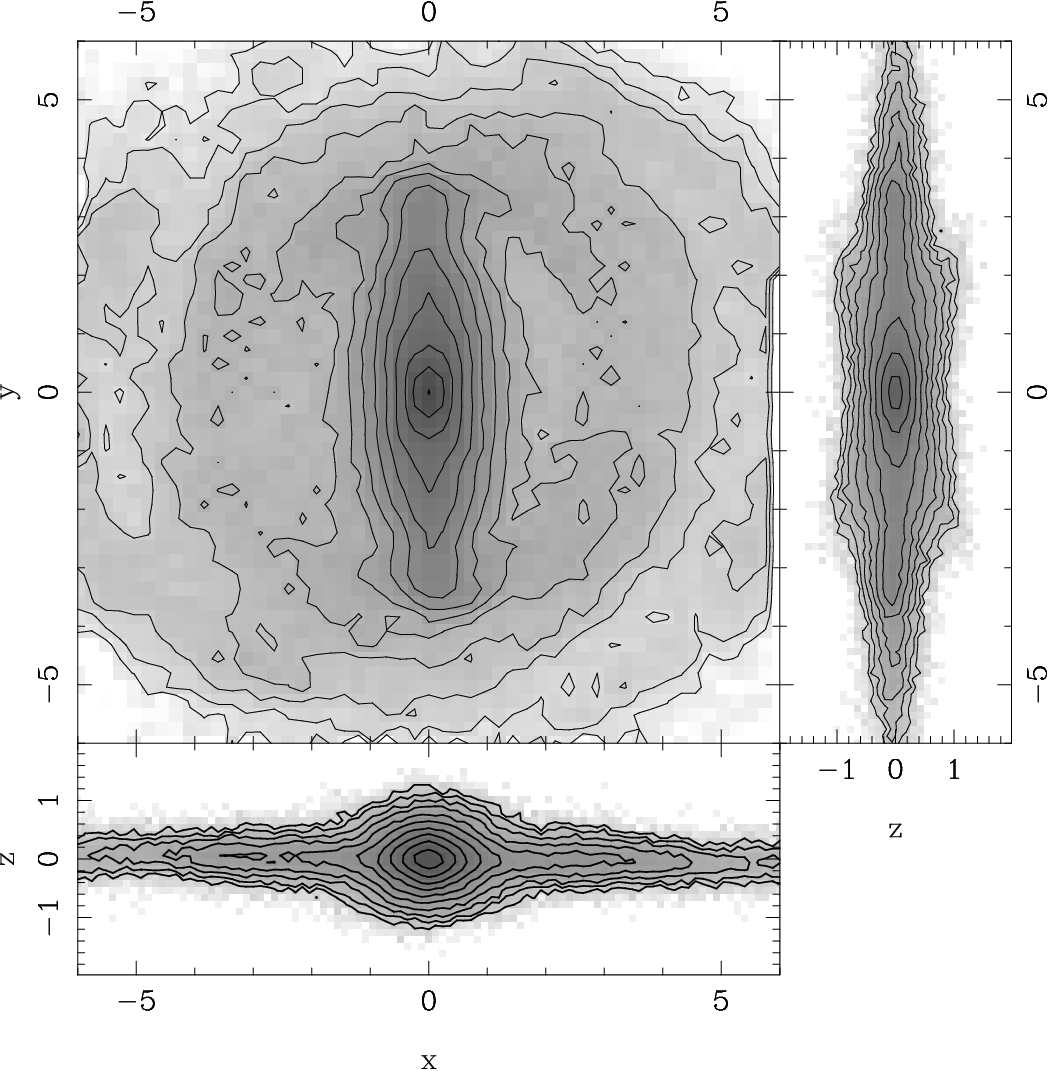}}
  \parbox[t]{30mm}{\vspace*{-50mm}\centerline{$\Delta T=0$}}
  \parbox[t]{70mm}{{}
    \includegraphics[scale=0.4,angle=0]{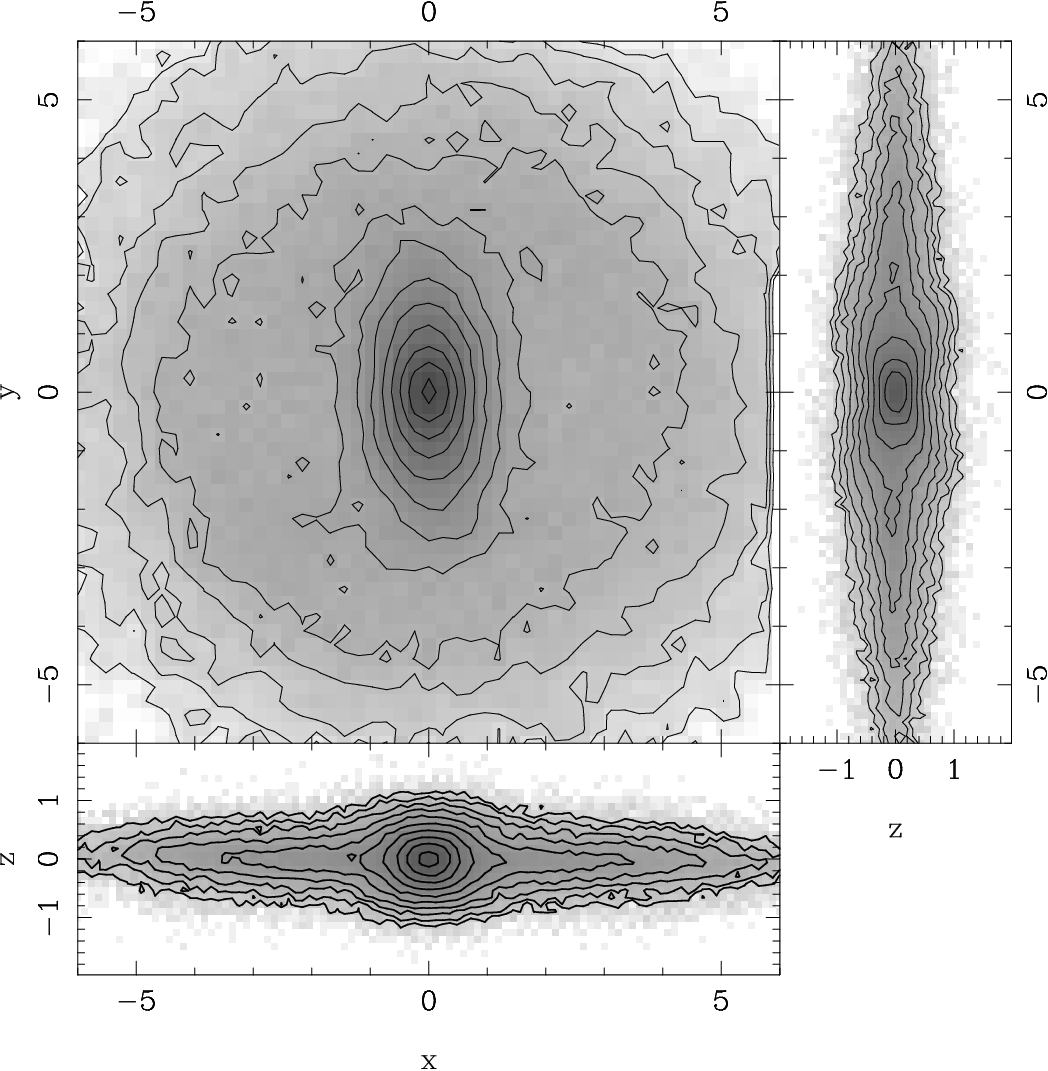}}\\[1ex]

  \parbox[t]{70mm}{{}
    \includegraphics[scale=0.4,angle=0]{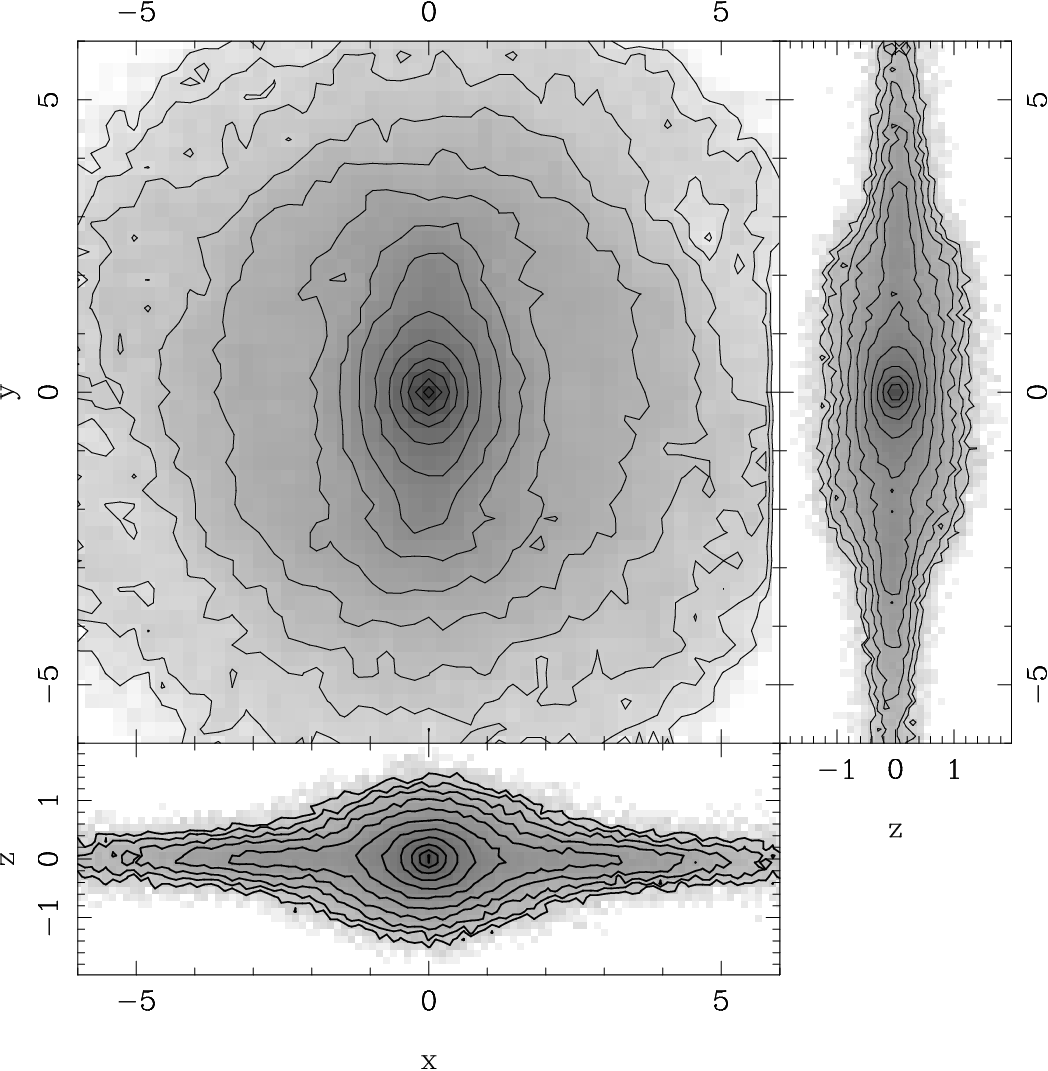}}
    \parbox[t]{30mm}{\vspace*{-50mm}
      \centerline{$\Delta T=400$}
      \centerline{$M_{\mathrm{CMC}}=0.05$} }
  \parbox[t]{70mm}{{}
    \includegraphics[scale=0.4,angle=0]{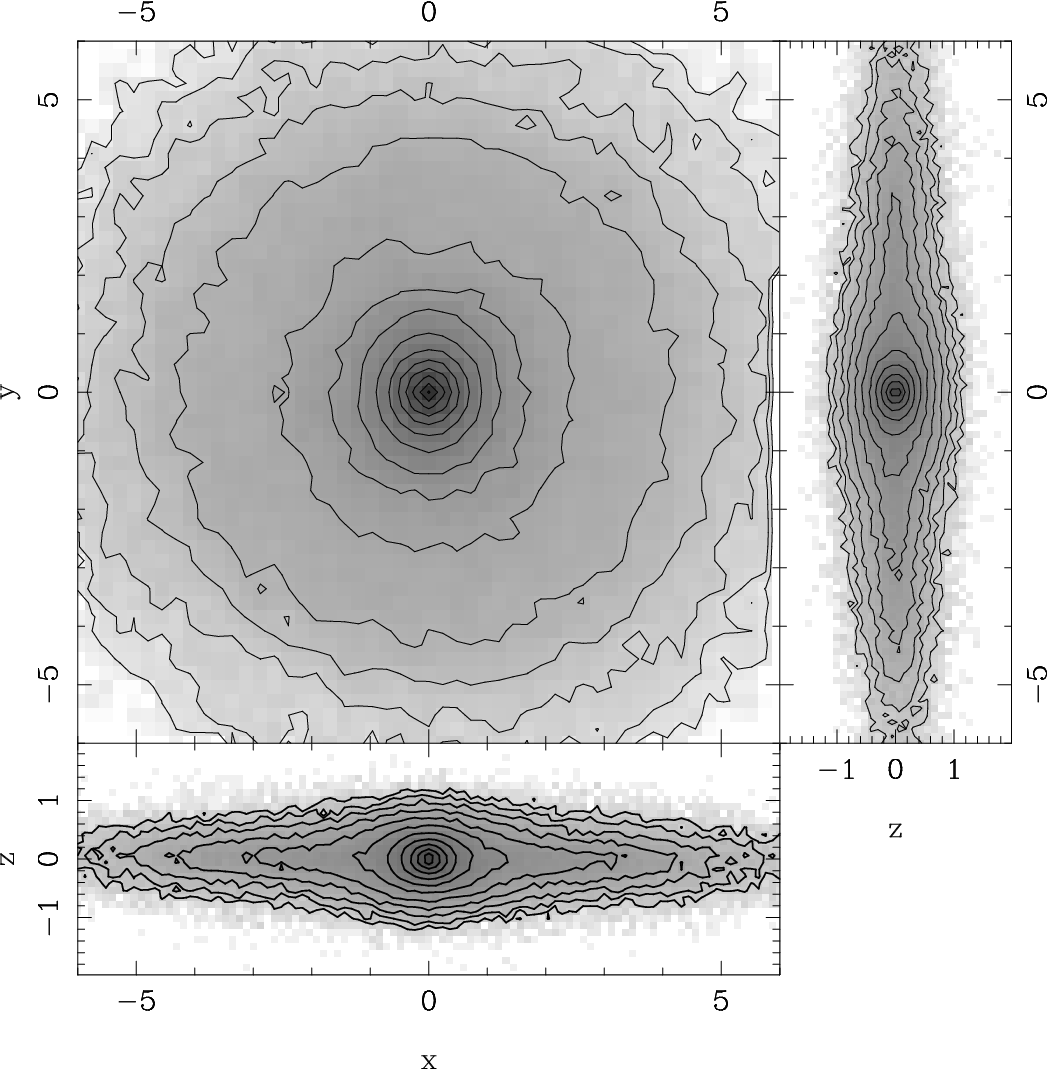}}\\[1ex]

  \parbox[t]{70mm}{{}
    \includegraphics[scale=0.4,angle=0]{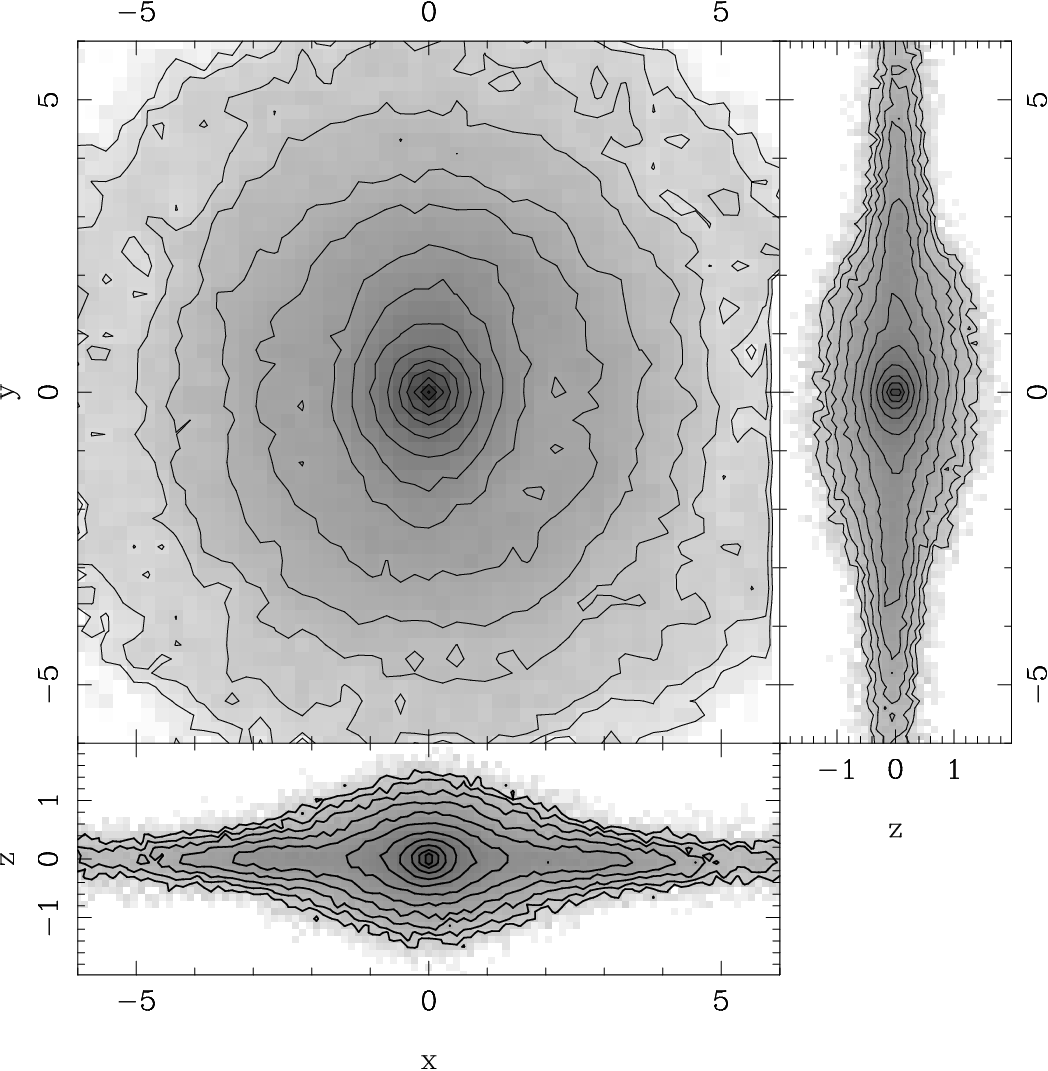}}
    \parbox[t]{30mm}{\vspace*{-50mm}
      \centerline{$\Delta T=400$}
      \centerline{$M_{\mathrm{CMC}}=0.1$} }
  \parbox[t]{70mm}{{}
    \includegraphics[scale=0.4,angle=0]{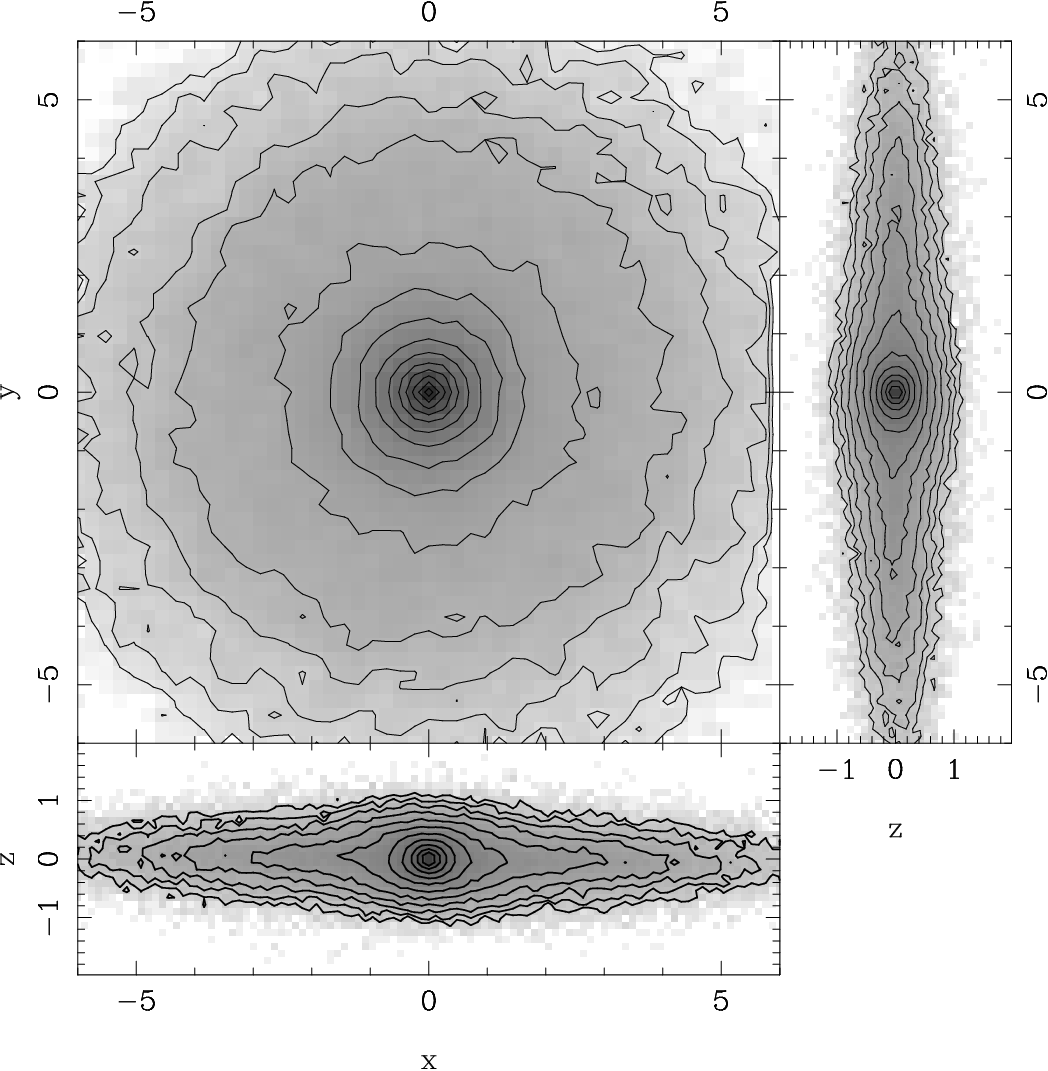}}
  \caption{Effect of a CMC on a MH-type (\emph{left}) and 
    MD-type (\emph{right}) model. The upper panels show the disc
    component at the time the CMC is introduced and the others at
    $\Delta T=$ 400 later. The mass of the CMC is 0.05 for the middle
    panels and 0.1 for the lower two, while $r_{\mathrm{CMC}}=0.01$ in
    both cases. Each sub-panel shows one of the three orthogonal
    views of the disc component. The right sub-panel gives the edge-on
    side-on view, the lower sub-panel gives the edge-on end-on view
    and the main sub-panel the face-on view. The projected density of
    the disc is given by grey-scale and also by isocontours (spaced
    logarithmically).}
  \label{fig:3v_MHMD}
\end{figure*}

\subsubsection{MH-type bars}
The effect of a CMC on a MH-type bar is shown in the left panels of
Fig.~\ref{fig:3v_MHMD}. The upper left panel shows the three views of
the disc component before the CMC is introduced. The face-on view
shows a strong bar with ansae at its ends, similar to what is seen in
many early types strongly barred galaxies \citep[good examples can be
seen on pages 42 and 43 of the Hubble Atlas,][]{Sandage1961}.  Seen
side-on, it shows a clear peanut, or mild `X', shape, whose extent is
somewhat shorter than the bar, as expected \citep[for a discussion
see][]{Athanassoula2005b}. Seen end-on, the bar can easily be mistaken
for a bulge and can only be distinguished from it by kinematical
measurements \citep{BureauAthanassoula2005, Athanassoula2005a,
  Athanassoula2005b}. 

The middle and lower panels on the left of Fig.~\ref{fig:3v_MHMD} show
the disc component a $\Delta T=400$ after the CMC has been introduced.
With the units proposed in section~\ref{subsec:initcond}, this is
equivalent to 5.6\,Gyrs. The mass of the CMC in these two cases is
0.05 and 0.1, respectively and $r_{\mathrm{CMC}}=0.01$. We note that
the smaller of the two masses did not destroy the bar, but made it
considerably less strong. In particular, the final bar is considerably
shorter. Since its width has not changed considerably, its axial ratio
(minor to major axis) has become considerably larger.  The heaviest of
the two CMCs has also not destroyed the bar, although its mass is as
much as 10\% of that of the disc. The resultant bar, however, is
rather short and oval. A CMC with a mass of 0.2 is sufficiently
massive to destroy the bar and make the disc near-axisymmetric (not
shown here).

\begin{figure*}
  \setlength{\unitlength}{2cm}
  \centerline{              \hspace{10mm}
    $M_{\mathrm{CMC}}=0$    \hspace{30mm}
    $M_{\mathrm{CMC}}=0.05$ \hspace{40mm}
    $M_{\mathrm{CMC}}=0.1$
  }

  \smallskip
  \includegraphics[scale=0.86,angle=0]{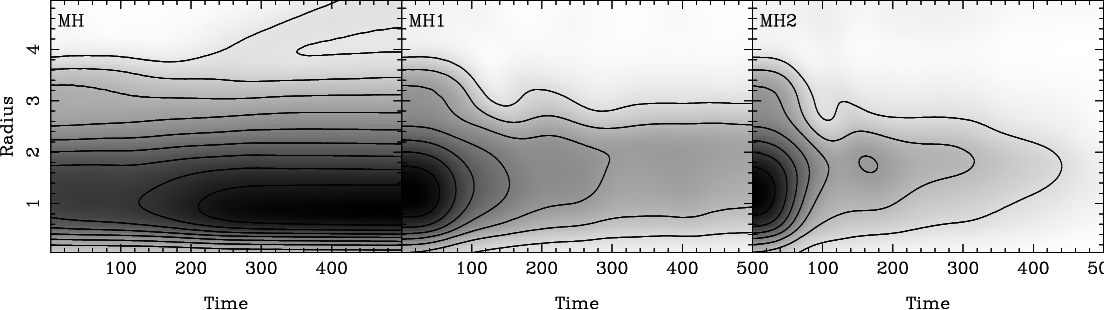}

  \smallskip
  \includegraphics[scale=0.86,angle=0]{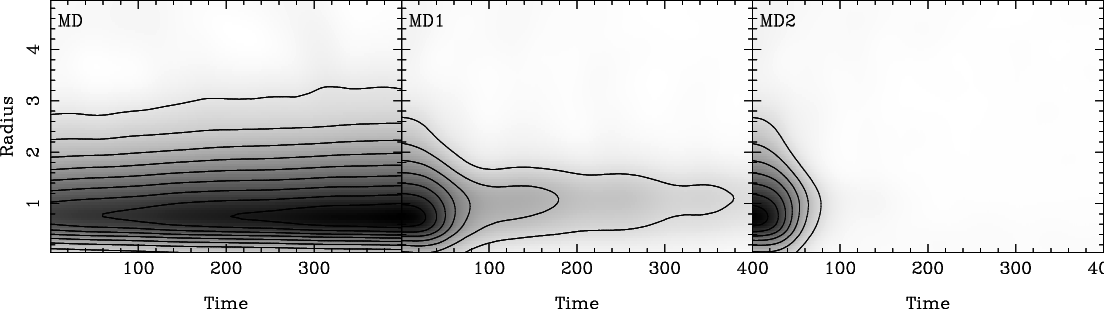}
  \caption{Bar amplitude ($m=2$ Fourier component of the amplitude) as a
    function of time and radius, for the three MH-type (\emph{top})
    and three MD-type (\emph{bottom}) simulations snapshots of which
    are shown in the left and right panels of Fig.~\ref{fig:3v_MHMD},
    respectively. The grey-scale and isocontour levels are set
    separately in each panel, so as to show best the relevant
    features. The isocontour levels are spaced linearly and darker
    shades correspond to higher amplitudes. The name of the simulation
    is given in the left corner of each panel.}
  \label{fig:2Damplitude_MHMD}
\end{figure*}

The CMC changes also the vertical structure of the bar. Before the
introduction of the CMC and seen side-on, the bar has a peanut-shape,
or mild `X'-form, with a maximum vertical extent at a distance from
the centre roughly equal to three quarters of the bar length and a
minimum in between the two maxima. The middle left panel,
corresponding to $M_{\mathrm{CMC}}=0.05$, show that the bar seen
side-on has a boxy shape, with a roughly constant vertical extent. The
radial extent of this boxy feature is roughly the same as that of the
peanut before the CMC was introduced. Thus, although the CMC brought a
considerable decrease of the bar length seen face-on, it brought
little, if any, decrease of the radial extent of the vertically
extended feature. The more massive CMC, $M_{\mathrm{CMC}}=0.1$, brings
a stronger change to the vertical structure. Seen side-on, the shape
is more elliptical than boxy. So the introduction of a CMC changes the
side-on outline from peanut to boxy, or, for very massive CMCs, to
elliptical. The radial extent of these features, however, does not
change much. In the end-on view also, the CMC makes changes in the
outline. The vertical extent increases in the central region and also
the bulge-like feature becomes considerably more extended radially.

\subsubsection{MD-type bars}
The right panels of Fig.~\ref{fig:3v_MHMD} are similar, but for an
initially MD-type bar. Before the CMC is introduced the bar is both
fatter and shorter, i.e.\ less strong than that of the previous
example, in agreement with the results of AM02. It also has no ansae.
Seen side-on, it has a boxy structure, but no peanut. The middle and
lower right panels show the disc component a $\Delta T=400$ after the
CMC has been introduced. The mass of the CMC in these two cases is
again 0.05 and 0.1, respectively and $r_{\mathrm{CMC}}=0.01$. We note
that the introduction of the CMC makes major changes. Seen face-on,
the bar has nearly disappeared even for $M_{\mathrm{CMC}}=0.05$, where
only close scrutiny allows one to see that the isophotes are not
circular, but slightly elongated in what used to be the bar region .
For $M_{\mathrm{CMC}}=0.1$, there is no trace of the bar anymore, the
isophotes being circular within the measuring errors.  Seen side-on,
the boxiness has more or less disappeared for $M_{\mathrm{CMC}}=0.05$,
and there is definitely no trace left by $M_{\mathrm{CMC}}=0.1$. Thus,
judging from the morphology of the three orthogonal views, the models
with $M_{\mathrm{CMC}}=0.1$ or 0.05, qualify as an SA galaxy, i.e. a
galaxy with no bar. Thus, even moderately light CMC can destroy the
bar in MD-type models. This is contrary to what was found above for
MH-type models and constitutes an important difference between the
two.

To check whether a less massive CMC can still destroy the bar we ran
four more simulations with $M_{\mathrm{CMC}}$ equal to 0.04, 0.03,
0.02 and 0.01, respectively. In all four cases the bar is clearly
visible by the end of the simulation, although considerably weakened.
We also ran simulations with $M_{\mathrm{CMC}}$ = 0.1 and
$r_{\mathrm{CMC}} > 0.01$ to check whether the bar is destroyed even
for CMCs with such large radii. We found that $r_{\mathrm{CMC}}$ has
to be at least as large as 0.1 for the bar not to be destroyed.

\subsection{The bar strength}
\label{subsec:bstrength}
\subsubsection{The bar strength as a function of time and radius}
\label{subsubsec:bstrengthtr}
Figure~\ref{fig:2Damplitude_MHMD} gives, for six simulations, the
amplitude of the $m=2$ (bisymmetric) component of the mass/density
distribution as a function of both time and radius. This is a simple
measure of the bar strength. The results for higher $m$ values will be
discussed in the next subsection. The left panels correspond to two
reference simulations in which no CMC has been introduced. They show
clearly a secular evolution, during which the bar gets slowly and
steadily stronger. We note that the maximum amplitude increases, but
also that both the innermost and the outermost parts become gradually
more bisymmetric.

\begin{figure*}
  \setlength{\unitlength}{2cm}
  \includegraphics[scale=0.9,angle=0]{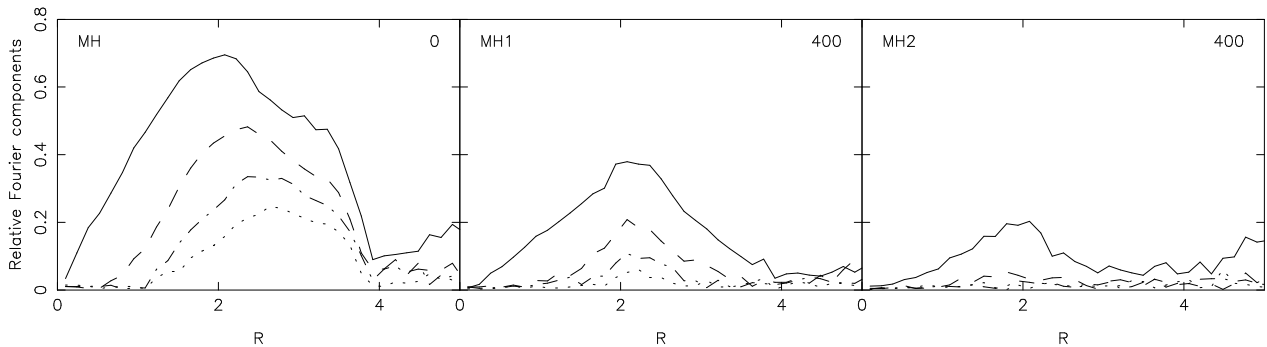}
  \caption{Relative amplitude of the $m=2$ (solid line), 4 (dashed
    line), 6 (dot-dashed line) and 8 (dotted line) Fourier components
    of the mass/density. The name of the simulation is given in the
    upper left corner and the time since the introduction of the CMC
    in the upper right corner of each panel.}
  \label{fig:mcomponMH}
\end{figure*}

\begin{figure*}
  \setlength{\unitlength}{2cm}
  \includegraphics[scale=0.9,angle=0]{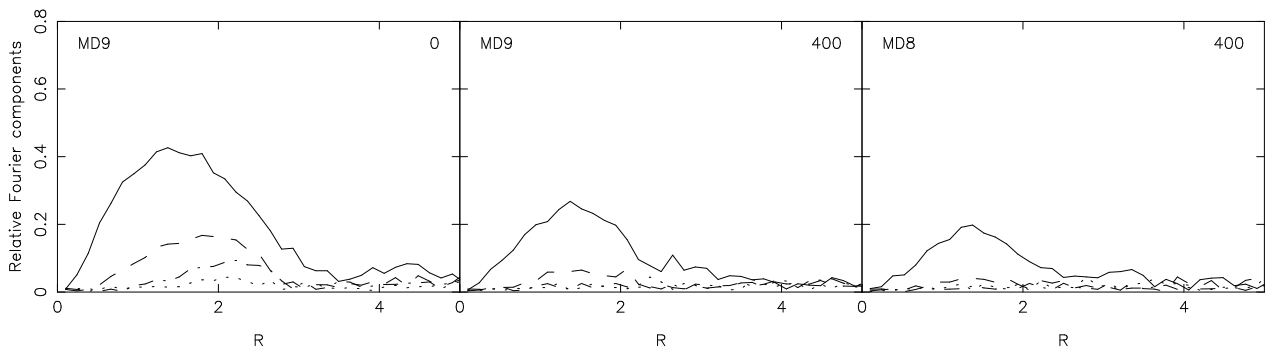}
  \caption{As in figure~\ref{fig:mcomponMH}, but for MD-type models.}
  \label{fig:mcomponMD}
\end{figure*}

The picture is different for simulations with a CMC. At early times,
up to 20 or 30 time units after the CMC has been introduced, the
evolution continues more or less as if no CMC had been introduced.
This is presumably because the CMC has to reach a limiting mass before
its effect can be felt by the bar and/or because the orbits in the bar
region take some time to adjust.  This time interval is longer in the
case of less massive CMCs. After that starts a second time interval,
which lasts roughly till the time the CMC reaches its maximum mass.
During this second phase the amplitude drops abruptly with time. A
third time interval starts after that and lasts till the end of the
simulation. Unless the bar has totally disappeared after the end of
the second interval, its amplitude continues to decrease in this third
phase, but at a considerably slower rate than in the second one. The
distinction between the second and the third time interval is sharper
for MD-type than for MH-type models and for more massive CMCs.
Figure~\ref{fig:2Damplitude_MHMD} shows clearly that the maximum
amplitude decreases steadily with time. But also that both the
innermost and the outermost parts of the disc become gradually more
axisymmetric. This is in agreement with what is seen in
figure~\ref{fig:3v_MHMD}. Indeed, it was clear there that at $\Delta
T=400$ after the CMC was introduced the bar has become shorter and
also that the innermost isophotes have become rounder. The
corresponding time evolution can be followed in
figure~\ref{fig:2Damplitude_MHMD}.

Comparison of the upper and lower panels of
Fig.~\ref{fig:2Damplitude_MHMD} shows that a given CMC can destroy
MD-type bars more efficiently than MH-type ones. Also comparison of
the central and right rows of panels shows that more massive CMCs are
more efficient for bar destruction, as expected.
 
\subsubsection{Higher $m$ values}
\label{subsubsec:m}

Fig.~\ref{fig:mcomponMH} shows the amplitude of the relative $m$ = 2,
4, 6, and 8 Fourier components of the mass/density for MH-type
simulations. They have been 
calculated as described in AM02. Before the introduction of the CMC
(left panel) all components have high amplitudes. In particular, their
maxima are 0.7, 0.48, 0.33 and 0.25, for $m$ = 2, 4, 6 and 8,
respectively. The introduction of the CMC brings a considerable
decrease of all these values. Thus, for a $M_{\mathrm{CMC}}$ = 0.05
and $\Delta T$ = 400 the maxima of the amplitudes for the various $m$s
fall to 0.54 ($m$ = 2), 0.43 ($m$ = 4), 0.32 ($m$ = 6) and 0.25 ($m$ =
8) of the initial maximum values, respectively. For $M_{\mathrm{CMC}}$
= 0.1 and the same time lapse the $m$ = 2 amplitude falls to 0.3 of
its initial value, while the other components drop to the noise level.

Fig.~\ref{fig:mcomponMH} quantifies the effect of the CMC and, more
importantly, shows that it has a stronger effect on the higher $m$
components. The latter can be understood as follows: As can be seen
from the left panel of Fig.~\ref{fig:mcomponMH}, before the CMC is
introduced the radius at which the relative amplitude has its maximum
increases with $m$ and comes closer to the outer regions of the bar.
These regions, as we saw in section \ref{subsec:existence} and
\ref{subsubsec:bstrengthtr}, are affected first and tend faster to
axisymmetry. In this way, the higher $m$ components will be affected
before the lower ones. This argument is further re-enforced when we
look at the positions of the maxima. For $m=2$ the maximum is at
roughly 2.1 and stays there after the CMC has been introduced, as can
be seen by comparing the left panel to the middle and right ones. On
the other hand the maximum of the $m=4$, and 6 components is roughly
at 2.4 at the time the CMC is introduced, and moves to 2.1, i.e. the
same radius as that of the $m=2$ maximum, 400 time units after a CMC
with $M_{\mathrm{CMC}}=0.05$ is introduced. Since the higher $m$
components are affected less than the lower $m$ ones, the iso-density
contours become fatter and/or less rectangular (and thus more
elliptical). Fig.~\ref{fig:3v_MHMD} confirms this.

Fig.~\ref{fig:mcomponMD} gives similar information, but for MD-type
simulations. As already discussed in AM02, MD-type models have lower
values of all the amplitudes and also lower values of the relative
amplitude of the higher $m$ components than MH-type models. Thus, for
MD-types, the CMC 
destroys the higher $m$ components even for relatively small values of the CMC
mass. This is illustrated in Fig.~\ref{fig:mcomponMD}, where we show
results for two
models with $M_{\mathrm{CMC}}$ equal to 0.01 and 0.02, respectively. 

\subsection{Velocity dispersions along the bar major axis}
\label{subsec:veldisp}

\begin{figure*}
  \setlength{\unitlength}{2cm}
  \includegraphics[scale=0.86,angle=0]{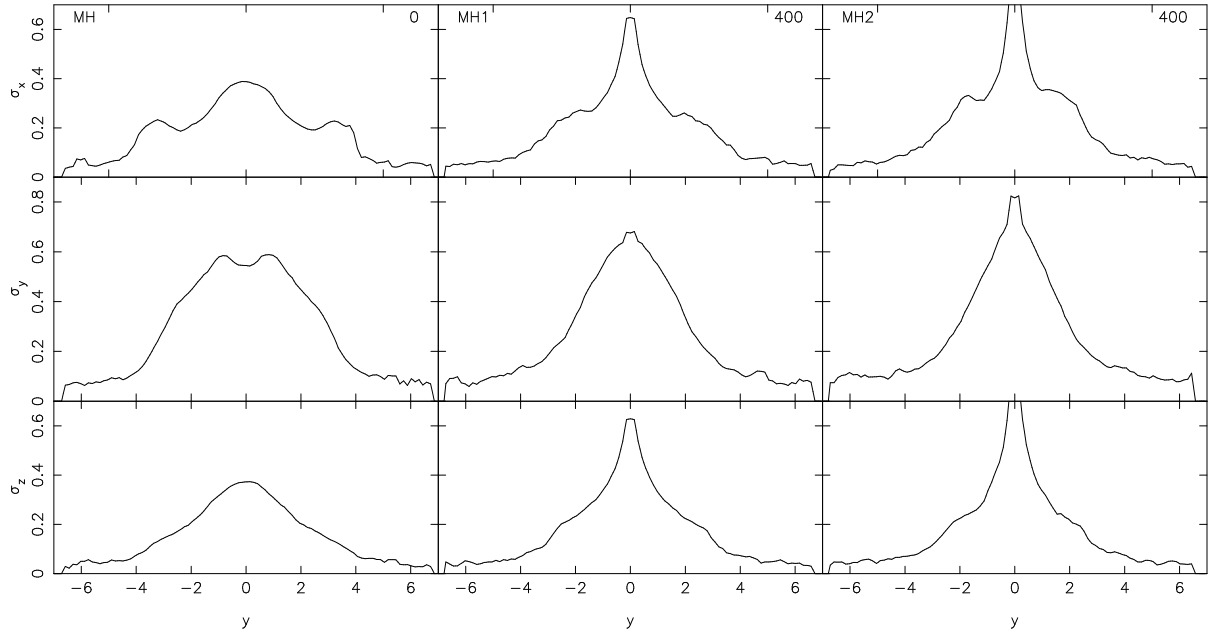}
  \caption{Velocity dispersion as a function of distance from the
    centre, as discussed in section~\ref{subsec:veldisp}. The left row
    of panels correspond to MH simulations at the time the CMC is
    introduced, while the central and right rows of panels correspond
    to simulations MH$_1$ and MH$_2$, respectively. The three
    components of the velocity dispersion, $\sigma_x$, $\sigma_y$ and
    $\sigma_z$, are given in the \emph{upper}, \emph{middle} and
    \emph{lower} rows of panels, respectively. The name of the
    simulation is given in the upper left corner and the time since
    the introduction of the CMC in the upper right corner of the upper
    panels. }
  \label{fig:sigma_MH}
\end{figure*}

Fig.~\ref{fig:sigma_MH} shows the three components of the velocity
dispersion as a function of distance from the centre. They have been
calculated as described in AM02. Namely, we isolate a thin strip of
particles centred on the bar major axis and having a width of 0.07.
The distance from the centre is calculated along this strip. In this
way, we avoid making an integration along the line of sight, since
this would impose the choice of a particular viewing angle. We also
bring out clearer the information on the motions of particles near the
bar major axis, which is of particular use when comparing with
periodic orbit structure. Nevertheless, any comparison with
observations can only be rough, since our presentation gives
information only on a subset of the orbits.

At the time the CMC is introduced, the velocity dispersions of model
MH show features characteristic of MH-type models (see AM02 for a more
complete description). The $\sigma_x$ component (perpendicular to the
bar) has a clear maximum at the centre and two secondary maxima on
either side of it. These can be due to orbits trapped around x$_1$
periodic orbits with loops at their apo-centres, and/or on the
superposition of orbits trapped around elliptical and around rectangular
shaped periodic orbits.  The CMC turns these maxima into plateaus and
moves them nearer to the centre. Both effects could be expected.
Indeed these maxima are only found for strong bars (AM02), while the
CMC has reduced the bar strength. Also these structures are linked to
the end of the bar and the CMC makes this shorter
(sections~\ref{subsec:existence} and \ref{subsubsec:bstrengthtr}). Of
course the clearest effect of the CMC is to raise very substantially
the central velocity dispersion, and more strongly so for the most
massive CMCs.
 
\begin{figure*}
  \setlength{\unitlength}{2cm}
  \includegraphics[scale=0.7,angle=-90.]{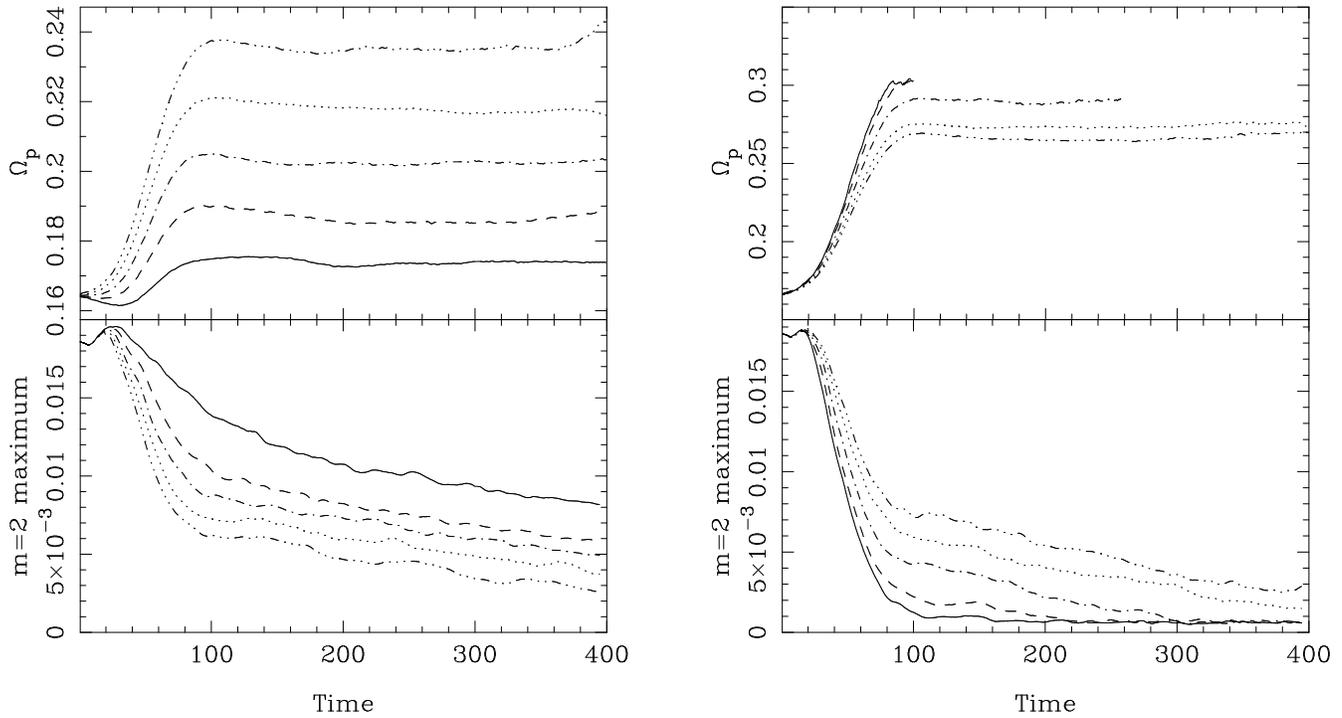}
  \caption{The upper panels give the pattern speed of the bar as a
    function of time. For some simulations and times the bar is not
    sufficiently strong to allow a correct measurement of the pattern
    speed, so we give no estimates. The lower panels give, for
    comparison, a measure of the bar strength, also as a function of
    time. As such a measure we use the maximum value of the amplitude
    of the $m$ = 2 component of the mass. The left panels refer to a
    series of simulations with different $M_{\mathrm{CMC}}$ : MD$_9$
    (solid line), MD$_8$ (dashed), MD$_7$ (dot-dashed), MD$_6$
    (dashed) and MD$_1$ (3 dots-dashed). The right panels correspond
    to a series of simulations with different $r_{\mathrm{CMC}}$ :
    MD$_{2}$ (solid line), MD$_{13}$ (dashed), MD$_{12}$ (dot-dashed),
    MD$_{11}$ (dashed) and MD$_{10}$ (3 dots-dashed).}
  \label{fig:omegap_Sb_MD}
\end{figure*}

At the time the CMC is introduced, the $\sigma_y$ component has a
small, but characteristic, minimum at the centre, surrounded closely
by two maxima on either side. Since the $\sigma_y$ is the component
parallel to the bar major axis, this feature should be mainly seen in
edge-on barred galaxies which are viewed nearer to end-on than to
side-on.  The CMC obliterates this feature and creates, instead, a
clear central maximum, as for $\sigma_x$. The $\sigma_x$ and
$\sigma_y$ components have quite different values at the centre in the
case with no CMC. As expected, however, the velocity distribution in
the central regions becomes more isotropic in the cases with CMC, and
more strongly so for the most massive CMCs.

The MD-type models show no clear characteristic features on their
velocity dispersion profiles (see fig. 13 of AM02). Thus the CMC only
introduces a sharp maximum at the centre, clear in all three
components of the velocity dispersion, as was seen already in the
MH-type models.

\subsection{CMC radius and growth time}
\label{subsec:misc}

We ran two simulations with smaller values of $r_{\mathrm{CMC}}$ (see
Table~\ref{tab:initcond_CMC}) and found that more centrally
concentrated CMCs are more efficient for bar destruction, thus
confirming a result already found by \cite{HasanPfennigerNorman1993},
with orbital calculations, and by SS, with simulations.

We also ran two simulations with a larger value of $t_{\mathrm{grow}}$
(see Table~\ref{tab:initcond_CMC}) and confirmed a result already
found by SS and by HH, namely that the value of $t_{\mathrm{grow}}$
does not influence much the final bar weakening. It only influences
the duration of the initial strong decrease. The transition between
the initial strong decrease and the later milder one is sharper in
simulations with larger $M_{\mathrm{CMC}}$ and/or smaller
$r_{\mathrm{CMC}}$ (see bottom panels Fig.~\ref{fig:omegap_Sb_MD}).

\subsection{Pattern speed}
\label{subsec:omegap}
\begin{figure}
  \setlength{\unitlength}{2cm}
  \includegraphics[scale=0.7,angle=-90.]{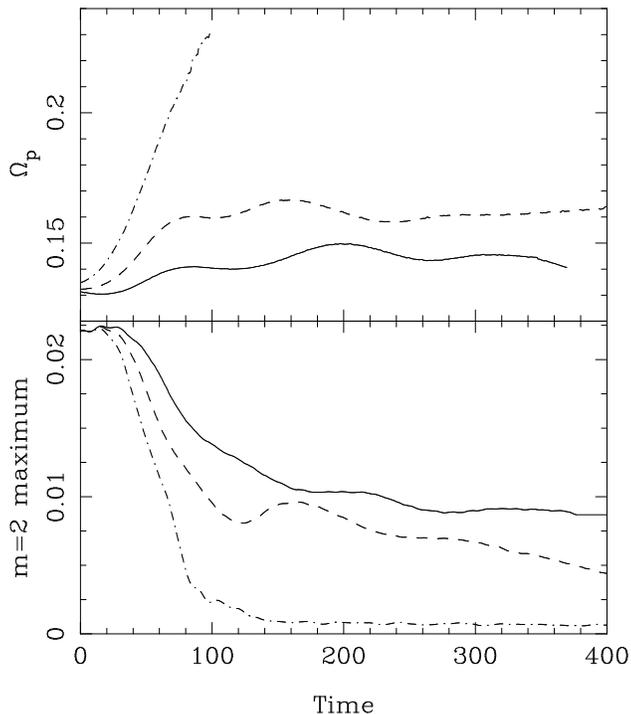}
  \caption{As for Fig.~\ref{fig:omegap_Sb_MD}, but now for MH-type
    models. Results are given for three simulations with different
    $M_{\mathrm{CMC}}$ : MH$_1$ (solid line), MH$_2$ (dashed) and
    MH$_3$ (dot-dashed). }
  \label{fig:omegap_Sb_MH}
\end{figure}

NSH mentioned that the introduction of the CMC results in an increase
of the bar pattern speed. In order to assess further the effect of the
CMC on the bar pattern speed, we measured this quantity in our
simulations. This was not always easy, since in many cases the bar
amplitude is so weak that the measurement was either not possible, or
gave a very uncertain value.  In most cases, some smoothing was
necessary. Nevertheless, from the cases where the measurement is
sufficiently precise, we can clearly see that the pattern speed \emph{
  increases} with time, as the bar gets weaker. This is shown, for two
sets of MD-type simulations, in Fig.~\ref{fig:omegap_Sb_MD} and for
MH-type simulations in Fig.~\ref{fig:omegap_Sb_MH}. For comparison, we
also plot a measure of the bar strength as a function of time. As such
we have chosen the maximum value of the amplitude of the $m$ = 2
component of the mass.  Results found with other measures of the bar
strength are similar.

The left panels of Fig.~\ref{fig:omegap_Sb_MD} give results from a series
of simulations with different values of the mass of the CMC, namely
simulations MD$_9$ ($M_{\mathrm{CMC}}$ = 0.01), MD$_8$ (0.02), MD$_7$
(0.03), MD$_6$ (0.04) and MD$_1$ (0.05). In the very first 20 time
units or so, the bar strength increases, as in simulations with no
CMC. During this time, at least in simulations with small CMCs, the
pattern speed decreases, again as in simulations with no CMC. This is
presumably due to the fact that the CMC has not grown enough to
influence the evolution sufficiently to reverse the trend. This
reversal indeed occurs at somewhat later times. We note that the bar
speed increases noticeably up to time 100, while the CMC mass
increases and the bar strength decreases most strongly. At later times
there may still be an increase in the pattern speed, but it is very
small and the data are too noisy to assess it with any certainty.

The right panels Fig.~\ref{fig:omegap_Sb_MD} give similar results, now
from a series of simulations with different values of the radius of
the CMC, namely simulations MD$_2$ ($r_{\mathrm{CMC}}$ = 0.01),
MD$_{13}$ (0.02), MD$_{12}$ (0.05), MD$_{11}$ (0,08) and MD$_{10}$
(0.1). For the value of $M_{\mathrm{CMC}}$ used in this sequence
($M_{\mathrm{CMC}}$ = 0.01) and the smallest of our
$r_{\mathrm{CMC}}$, the bar is practically destroyed by the time the
mass of the CMC has reached its final value. Thus CMCs with a yet
smaller radius give the same result. The very short-lived initial
increase of the bar strength is less clear here than in the previous
set of simulations. The basic result, however, stays the same. 

Fig.~\ref{fig:omegap_Sb_MH} shows similar plots, but now for MH-type
simulations. The results are similar, i.e. there is an increase in the
pattern speed during the time the strength of the bar
decreases sharply. This increase is quite noticeable in cases where the
decrease of the bar strength is important, and considerably less so in
cases with a milder decrease of the bar strength (as MD$_9$ or MH$_1$).

\section{Summary and Discussion}
\label{sec:discussion}
\subsection{Bar survival and destruction}
\label{subsec:survive}
In this paper we studied the effect of a central mass concentration
(CMC) on the bar that harbours it. We find that the CMC leads to a
decrease of the bar strength, which in some cases can be sufficiently
important to lead to bar destruction. More massive and/or more
concentrated CMCs are more efficient. The effect of the CMC depends
also on the type of the bar model. Strong bars, which we call MH-type,
because they form in simulations with massive haloes and which are
morphologically similar to \citeauthor{ElmegreenElmegreen1985}'s
(\citeyear{ElmegreenElmegreen1985}) flat-profile bars, are less prone
to destruction than weaker bars, which we call MD-type, because they
form in simulations with massive discs and which are morphologically
similar to \citeauthor{ElmegreenElmegreen1985}'s exponential-profile
bars.

The effect of the CMC is strongest in the innermost and the outermost
parts of the bar, where it makes the density distribution more
axisymmetric. This leads to shorter and fatter bars, whose iso-density
contours are more elliptical than rectangular. The CMC also affects
the pattern speed of the bar, which grows noticeably while the mass of
the CMC grows and the strength of the bar decreases strongly.  This is
in good agreement with the anti-correlation between bar strength and
pattern speed found in A03 (see figures 16 and 17 in that paper). The CMC also
influences the velocity dispersions. In particular, after the
introduction of the CMC, the velocity dispersion profiles along the
bar major axis show a strong central maximum, while the secondary
maxima on either side of the centre, seen in the side-on projection of
strong bars, become less strong and approach the centre.

In order to understand why a given CMC can destroy a bar in some
cases, while only weakening it in others, it is necessary to study the
time evolution of the orbital structure in the two cases. For this, one
needs to study the
orbital structure at a sequence of times during the evolution by
calculating the potential and pattern speed at these times and using
the corresponding positions and velocities of the simulation particles  
(as done e.g. in A02 and A03). The evolution of
the orbital structure along this time sequence will give information
on the evolution of the orbital structure during the simulation and should
help explain the difference between the effects of a CMC on MH-type
and on MD-type bars. This analysis is beyond the scope of this
paper and will be the subject of a future paper. It is possible,
however, to make some remarks at this stage. A MH-type bar which
survived with a decrease of its amplitude was initially thinner, i.e.\ 
had a smaller ratio of minor to major axis, than a MD-type bar which
was destroyed. This means that the orbits in the former are more
eccentric than those in the latter \citep{Athanassoula1992a}, i.e.\ 
for a given orbital major axis, the particles in the former will come
nearer to the CMC than in the latter. Thus, one would expect that the
bar structure in the MH-type bar would be more affected than the bar
structure in the MD-type, and this is indeed what is found by
\cite{HasanNorman1990} in their orbital calculations. Furthermore, as
can be seen in Fig.~\ref{fig:faceon_prof}, the density distribution in
MH-type bars is more centrally concentrated than in MD-types, so that
the particles come closer to the CMC. Both effects suggest that
MH-type bars could be destroyed more easily, and yet fully self-consistent
simulations show the opposite.

It is not entirely clear what is wrong with this naive reasoning.
Possibly, the orbital structure of planar bars, as studied by
\citeauthor{HasanNorman1990} and HH, is a bad representation of the
orbits actually populated, which extend to non-negligible
heights. Indeed, \cite*{SkokosPatsisAthanassoula2002} have shown that
the backbone of the bar is formed by the 3D families of the x$_1$ tree, and not
the 2D x$_1$ family. Thus, the fact that
the planar orbits are unstable does not necessarily imply that the bar
has to be destroyed. Finally, the 3D orbital-structure study of
\cite{HasanPfennigerNorman1993} is restricted to a single bar model
and thus cannot shed any light on the differences between MH-type and
MD-type bars.

Additionally, the halo may play a role. Indeed, in galaxies with no
CMC, MH-type haloes can absorb much more angular momentum from the bar
than MD-type haloes, thus causing the bar to grow stronger in MH-type
galaxies (A03). A similar effect could also occur in disc galaxies
with a CMC. Namely, in galaxies with a CMC, the extra angular momentum
that can be absorbed by the MH-type haloes could well cause the bar to
decrease less in MH-type than in MD-type galaxies, which is indeed
what our simulations show. This alternative could also explain why SS
find only mild differences between their strong and their weak
bar. Indeed, if the difference is due to the halo, it would not show
in these simulations which have a rigid halo.

\subsection{Comparison with previous work}
\label{subsec:comp}
There is good qualitative agreement between the results of all
$N$-body studies so far, namely NSH
(\citealt*{NormanSellwoodHasan1996}), SS
(\citealt*{ShenSellwood2004}), HH (\citealt{HozumiHernquist2005}), and
our study. First and foremost, all find that the growing CMC
results in a decrease of the bar strength. The disagreement is more
quantitative than qualitative and concerns whether the decrease in bar
amplitude is sufficiently strong to destroy the bar. There are further
agreements, in that the pattern speed increases as the bar strength
decreases (NSH \& this study) and that more massive and/or more
centrally concentrated CMCs weaken the bar more than less concentrated
ones (all studies). All the above are also in agreement with
orbital-structure studies \citep{HasanNorman1990,
  HasanPfennigerNorman1993}.  Finally, there is agreement that
$t_{\mathrm{grow}}$ does not influence much the final bar weakening,
but only influences the rate at which the bar strength decreases.

Quantitative comparisons are much more difficult to make, mainly
because of the differences between the various studies, affecting (1)
the bar and (2) halo models (life or rigid), (3) the Poisson solvers
and time steps, (4) the dimensionality of the motion (2D or 3D), and
(5) the initial conditions.

\subsubsection[]{\cite{NormanSellwoodHasan1996} and 
               \cite{ShenSellwood2004}}
\label{subsubsec:NHS:SS}
Despite all these differences, NSH, SS, and our study all agree in
that a CMC mass of at least a few per cent of the disc mass is
required to destroy a bar. SS find that a CMC with
$M_{\mathrm{CMC}}=0.04$ and $r_{\mathrm{CMC}}=0.001$ are necessary to
destroy the bar, while for $M_{\mathrm{CMC}}=0.1$ and
$r_{\mathrm{CMC}}=0.1$ the bar is weakened but not destroyed. Our
fiducial $r_{\mathrm{CMC}}$ is between these values and, for one of
our bar models, the bar is destroyed and for the other it is
considerably weakened for $M_{\mathrm{CMC}}=0.05$ or 0.1. 

The most notable difference between these previous studies and ours
is the treatment of the halo (rigid vs.\ fully self-consistent), since
SS used a rigid halo for computational economy. In support of the
adequacy of this choice they run one simulation with a live halo and
compare it to a rigid halo case. Unfortunately the two runs are not
very similar initially, so that an assessment of the importance of the
halo response does not follow straightforwardly from a comparison.
Indeed, Fig.~8 in SS shows a difference in the initial bar amplitude
of the same order as the difference between their strong and weak bar
cases (their Fig.~5). Moreover, their comparison is done for a halo
whose density is low in the disc region, which will indeed restrict
the influence of the halo on the bar evolution. Simulations with a
dense halo would necessarily show a much stronger influence. In
particular, a rigid halo will not allow halo material to form a cusp
around the CMC. Furthermore, the importance of a live halo for the
correct description of bar evolution has been clearly demonstrated in
some cases by comparing rigid and live halo simulations (A02). For
these reasons, we have adopted a live halo in our simulations and
believe this to be crucial for a correct astrophysical understanding.

SS studied two bars of different strength and found only mild
differences. On the other hand, we find significant differences
between our MD-type and MH-type bars: the former is destroyed for an
$M_{\mathrm{CMC}}=0.1$ and an $r_{\mathrm{CMC}}=0.01$, while the
latter is only weakened. This is not a real disagreement, since the
difference between the weak and strong bar of SS is considerably less
than the difference between our MH-type and MD-type bars. However, if
the robustness of our MH-type bars originates (at least partly) from
the interaction with its halo, simulations like those of SS may not be
suitable for interpreting the evolution of galaxies with massive
haloes.

There are notable differences between the two sets of studies with
respect to their softening. We use a softening of 0.03 (i.e.  0.02
Plummer equivalent), a fiducial $r_{\mathrm{CMC}}$ of 0.01 and a range
of $r_{\mathrm{CMC}}$ values between 0.005 and 0.1. Thus our fiducial
$r_{\mathrm{CMC}}$ is half the softening, while in most of our adopted
range it is equal to or bigger than the softening. We also use a
softening kernel which decreases faster than the standard Plummer one,
thus limiting its effect at larger radii. To test further the effect
of the softening, we ran a series of simulations with larger
softening, namely 0.06 and 0.09 (i.e.\ equivalent Plummer softening
of 0.04 and 0.06) and found only very small differences, e.g.\ in the
bar strength. This argues strongly that our softening is adequate for
the problem at hand. SS have been bolder than us in the use of the
softening.  They use a softening of 0.02, for their strong bar case,
or 0.05 for their weak bar case, while studying the effect of CMCs
with $r_{\mathrm{CMC}}$ ranging between 0.0003 and 0.1, with a
fiducial value of 0.001. This makes their softening 20 to 50 times
larger than their fiducial $r_{\mathrm{CMC}}$, the ratio reaching, for
their most concentrated CMC, roughly 67 for their strong bar case and
167 for their weak bar.  In their figure 7, they compare three
simulations with softening 3, 7 and 17 times the CMC radius respectively
and find differences in the bar strength only of the order of, or
smaller than, 10\%.

\subsubsection[]{\cite{HozumiHernquist2005}}
\label{subsubsec:HH}
The recent study by HH differs notably from both previous studies (NSH
\& SS) and from ours. In particular, HH find that a CMC with 1\% or even
0.5\% of the disc mass is sufficient to destroy a bar (for
$r_{\mathrm{CMC}}=0.01$ disc scale lengths). It is not clear at this
stage what actually causes this discrepancy. HH blamed the different
initial conditions, in particular the fact that they have an
exponential disc, while NSH \& SS use a Kuzmin-Toomre disc. 
Our discs, however, are also initially exponential and yet our results
are in agreement with those of SS, but not with those of HH. We
discuss below some other possible explanations of the discrepancy.

One alternative is that there is a subtle numerical difference
between the two sets of simulations, in particular the resolution of
the SCF code and the time stepping. A poor resolution, corresponding
to large effective softening, should, however, decrease the mass of any
cusp forming around the CMC and hence reduce the destructive effect.
Furthermore, HH have tested the number of terms they use and found it
to be sufficient. An insufficiently small time step may entail an
artificial bar destruction. Both SS and this study use a variable time
step, which is very small in the vicinity of the CMC, contrary to HH,
who use a constant time step. In the same units as we use here, HH
have a $\Delta t=0.01$, or 0.005, depending on the model. This
contrasts with the time step which we have adopted, which, in the
vicinity of the CMC, is roughly 0.0005 or 0.0002. The time steps may
be compared with the dynamical time at $r\to0$ due to the CMC alone
\begin{equation}
  t_{\mathrm{dyn,\,CMC}}=2\pi\sqrt{r^3_{\mathrm{CMC}}/GM_{\mathrm{CMC}}}
\end{equation}
for a Plummer model (used by NSH, SS, and us) and $\sqrt{3/4}$ shorter
for the CMC model of HH. For the $M_{\mathrm{CMC}}=0.005$ model of HH,
we find $t_{\mathrm{dyn,\,CMC}}\approx0.077$. 

An obvious difference between HH on the one hand and SS and our work
on the other is that the former simulations are 2D, while the latter
are 3D. \cite{NormanSellwoodHasan1996}, however, compared 2D and 3D
simulations and found no important differences, capable of accounting
for the discrepancy we wish to explain. Given that
\citeauthor{NormanSellwoodHasan1996} consider only a single
bar model, it would be desirable to have more in order to better understand
the relation between 2D and 3D results.

\subsubsection{Simulations with gas}
\label{subsec:gas}
Gas flowing in a barred galaxy potential will form shocks along the
leading edges of the bar \citep{Athanassoula1992b}, leading to an
important inflow which can create a CMC. A number of $N$-body+SPH
calculations have followed this evolution and found that the CMC that
is thus formed can destroy the bar \citep{FriedliBenz1993,
  Friedli1994, BerentzenEtal1998}. \citeauthor{Friedli1994} found
that, when the gas is sufficiently concentrated so as to form a CMC of
2\% of the disc mass, the bar is destroyed in about 1 Gyr.
\citeauthor{BerentzenEtal1998} found a very similar result when the
CMC formed by the inflow had a mass equal to 1.6\% that of the galaxy
within 10 kpc. and a characteristic radius of the order of a few 100
pcs. The bar is destroyed also in the simulations of
\cite{BournaudCombes2002}. The two first studies use SPH
\citep{Monaghan1992} to model the gas, while the third one uses a
sticky particles algorithm \citep{Schwarz1981}.

The CMC mass and radius estimates obtained from such simulations
cannot be compared with those obtained by purely $N$-body simulations.
Of course, in both cases the growth of the CMC makes a large fraction
of the x$_1$ orbits unstable, and this family is known to be the
backbone of the bar \citep{ContopoulosPapayannopoulos1980,
  AthanassoulaEtal1983}. In cases including gas, however, there is an
extra effect, absent from the purely stellar cases. It is now known
that the growth of the bar is governed by the angular momentum
exchanged between the inner parts of the disc (bar) and the halo plus
outer disc (A03). In cases with gas, there is an extra component
taking part in this exchange process.  \cite{BerentzenEtal2004} have
shown that the gas, by giving angular momentum to the inner disc, will
damp the bar growth.  Thus the presence of a gaseous component should
favour bar destruction so that the CMC mass necessary to achieve this
should be smaller in cases with a sizeable gaseous component than in
purely stellar ones.

\subsection{Comparison to observations}
\label{subsec:observations}
As noted in the introduction, the mass of observed super-massive black
holes is of the order of $10^{-3}$ times the mass of the bulge that
harbours them. Unfortunately, neither our simulations, nor those of
NSH\,\&\,SS and HH, include any case with a live bulge. Bulges,
however, are in general less massive than the corresponding discs, so
the value of $10^{-3}M_{\mathrm{d}}$ can be considered as an upper
limit of the super-massive black hole mass. This is two orders of
magnitude lower than our fiducial value and one order of magnitude
smaller than the smallest value considered here. Making comparisons
with the CMC radius is less straightforward. Following previous
studies, we assume that our results are relevant to super-massive
black holes since the CMC radius is much smaller than its influence
radius; more than an order of magnitude for our fiducial cases.
Therefore, in as much as our results can be applied to super-massive
black holes, they argue that these cannot destroy the bar that
harbours them.

Observations show that important mass concentrations, sometimes as
large as $10^9 M_{\odot}$, can often be found in the central parts of
strongly barred galaxies \citep{SakamotoEtal1999, ReganEtal2001,
  HelferEtal2003, KormendyKennicutt2004}. These are either in the form
of molecular gas discs, or of discy bulges \citep[for a description
see][]{Athanassoula2005b}. Simulations that do not include gas
give only partial information on the effect of such a CMC on the bar,
since they can not evaluate the role of the gas in the angular
momentum exchange. In other words, they can give information on the
effect of the CMC as such, but not on the effect of its formation. The
following discussion will therefore, by necessity, neglect the latter
effect. Assuming a 
characteristic disc mass of 5 $\times10^{10}\,M_{\odot}$, we get a ratio
$M_{\mathrm{CMC}}/M_{\mathrm{d}}=0.02$, i.e. five times smaller than
our fiducial $M_{\mathrm{CMC}}$. The sizes of these CMCs are between
0.1 and 2 kpc, i.e. between 0.03 and 0.6 times the exponential disc
scale length of the Milky Way.  Assuming that the characteristic scale
length is half the extent, we find that the corresponding values for
the $r_{\mathrm{CMC}}$, in the units used here, are between 0.015 and
0.3.  Since in this paper we have used values of $r_{\mathrm{CMC}}$
between 0.005 and 0.02, the observed gaseous CMC will, for a same
mass, be less destructive than the CMCs used here. 
We thus come to the conclusion that such CMCs, seen their masses
and not withstanding the effect of their formation, are not capable of
destroying bars.

\cite{SakamotoEtal1999} and \cite{ShethEtal2005} show that the degree of
gas concentration in 
the central kpc is higher in barred than in unbarred galaxies. This
would not be possible if the molecular gas CMCs could destroy a bar.
It is, however, in good agreement with our results. The torques from a
bar can push gas to the centre and, for a given gas reservoir, the
stronger the bar the more important this central gaseous concentration
will be \citep{Athanassoula1992b}. It is thus reasonable, if the CMC
does not destroy the bar, to expect that stronger gas concentrations
will be found in barred rather than in unbarred galaxies, as is indeed
borne out by the observations of \citeauthor{SakamotoEtal1999} and
\citeauthor{ShethEtal2005}. 
 
\section*{Acknowledgements}
We thank A.~Bosma and S.~Hozumi for stimulating discussions and an
anonymous referee for comments that improved the presentation. E.A. and
J.C.L. thank the INSU/CNRS, the region PACA and the University of
Aix-Marseille I for funds to develop the computing facilities used for
the calculations in this paper.

\label{lastpage}

\end{document}